\documentclass[aps,pra,floatfix,twocolumn,amsmath,amssymb,superscriptaddress]{revtex4}
\usepackage{times}
\usepackage{latexsym}
\usepackage{graphicx}
\usepackage{verbatim,times,bbm}
\usepackage{color}
\usepackage{appendix}

\usepackage{float}
\usepackage[caption = false]{subfig}

\usepackage{color}

\begin{document}

\title{Thermal effects on coherence and excitation transfer}
\author{Laleh Memarzadeh\footnote{Corresponding author. email: memarzadeh@sharif.edu}}
\affiliation{Department of Physics, Sharif
University of Technology, Teheran, Iran}

\author{Azam Mani\footnote{email: {mani.azam@ut.ac.ir}}}
\affiliation{Department of Engineering Science, College of Engineering, University of Tehran, Iran}

\begin{abstract}
To control and utilize quantum features in small scale for practical applications such as quantum transport, it is crucial to gain deep understanding of quantum characteristics of states such as coherence. Here by introducing a technique that simplifies solving the dynamical equation, we study the dynamics of coherence in a system of qubits interacting with each other through a common bath at non zero temperature. Our results demonstrate that depending on initial state, environment temperature affect coherence and excitation transfer in different ways. We show that when initial state is incoherent, as time goes on, coherence and probability of excitation transfer increase. But for coherent initial state, we find a critical value of temperature, below which system loses its coherence in time which diminishes the probability of excitation transfer. Hence in order to achieve higher value of coherence and also higher probability of excitation transfer, temperature of the bath should go beyond that critical value. Stationary coherence and probability of finding excited qubits in steady state, are discussed. We also elaborate on dependence of critical value of bath temperature on system size.     
\end{abstract}


\date{\today}

\maketitle

\section{Introduction}
Recently lots of attention has been devoted to quantum coherence, as success of many quantum algorithms and quantum information processing tasks, is relied on this quantum mechanical feature. Apart from its impact in various quantum information tasks, quantum coherence plays significant role in different areas of research such as solid state physics \cite{solid,Nori}, spin models \cite{KarpatFanchini,ÇakmakFanchini,Vianna}, quantum thermodynamics \cite{Popescu,Rudolph,Gour,Oppenheim} and  biological systems \cite{Lloyd,Grondelle,HuelgaPlenio,Lambert}. Developments based on this quantum property which does not have any counterpart in classical realm, led to consider it as quantum resource \cite{Aberg,Baumgratz}. Different measures for quantifying quantum coherence have been introduced \cite{Aberg,Baumgratz,Girolami, Shao,Streltsov, Yuan,Pires} and resource theory of quantum coherence is well established \cite{Winter,Lewenstein}.\\  
\\
Beside formulation and interpretations of quantifying measures for coherence, dynamics of coherence under noise is of interest. For practical applications it should be taken into account that unavoidable interaction of system with its surrounding environment induces noise on the system. In fact dealing with noise effects is one of the main obstacles in implementation of quantum information tasks. As long as information contents of systems is of importance, error correcting codes or error correction based on feedback \cite{KnillLaflamme,WisemanMilburn,GregorattiWerner,MMM} are powerful tools to combat noise effects. For resources such as entanglement and coherence, analysing their dynamics under noise is crucial as it gives us insight for designing successful experimental set ups and also propose models for different phenomena. For coherence most of work done in relation to noise, focus either on classifying state regarding their coherent properties or characterizing cohering, de cohering evolutions \cite{Pati,Mani,Adesso}. Furthermore, role of temperature of bosoic bath on dynamics of single qubit coherence is discussed in \cite{WuXu}. It is also shown that system-environemnt coherence can be generated when single qubit experience different types of noise \cite{Pozzobom}.\\
\\
Here we are interested in dynamics of coherence in a system with arbitrary number of qubits when qubits interact with each other through a common bath at non zero temperature. This model provides the ground to analyse coherence dynamics in presence of natural type of noise, discuss coherence properties of large number of particles or macroscopic coherence and also see the impact of bath temperature on this quantity. We use the framework of open quantum systems for our analysis. First we develop a technique for simplifying the complexity of solving master equations in Markovian dynamics by determining the invariant dynamical subspace regarding the symmetry and conserved quantities. Equipped by this technique we can answer important questions such as how dynamics of coherence and also having an excitation at one site, depend on initial conditions and what is the effect of bath temperature in such quantities. By rigorous analysis of coherence dynamics and probability that a qubit gets excited, we show that systems with higher value of initial coherence are not necessarily better for excitation transfer and achieving larger value of coherence in steady state. Our analysis can be extended for modelling excitation transfer in low dimensional systems such as quantum dots diluted in an isolating environment \cite{D'Ortenzi,Yamamoto} and can give insight about the role of coherence in excitation transfer in some biological systems \cite{Lloyd,Grondelle,HuelgaPlenio,Lambert}.\\
\\
Structure of the paper is as follows: In section \ref{sec:mod} we introduce the model. Section \ref{method-sec} is devoted to introducing a technique that simplifies solving the master equation. Two different and important initial states namely coherent and incoherent initial states are discussed in sections \ref{sec:Incoherent} and \ref{sec:Coherent} respectively. Conclusions will be drawn in section \ref{sec:Conclusion}.\\
\\
\section{Model}\label{sec:mod}
We consider a system of $N$ qubits embedded in a common bath. Hamiltonian of system is given by
\begin{equation*}
H_s=\frac{1}{2}\hbar\omega\sum_{i=1}^N\sigma_i^z,
\end{equation*}
where $\sigma_i^z$ is Pauli operator corresponding to spin operator along $z$-direction of qubit $i$. We denote the eigenstate of
$\sigma_i^z$ corresponding to $-1$ and $+1$ respectively with $|g\rangle_i$  (ground state) and $|e\rangle_i$ (excited state). Hence
Hilbert space $\mathcal{H}$ of $N$ qubits with dimension $d=2^N$ is spanned by orthonormal basis
${\{|g\rangle_i, |e\rangle_i\}}^{\otimes n}$.
Environment is a free bosonic field in thermal state $\rho_B=\frac{1}{Z_B}e^{-\beta H_B}$ ($Z_B$ is partition function of bath and $\beta=\frac{1}{KT}$, where $K$ is Boltzmann constant and $T$ is the temperature of bath) with the following Hamiltonian
\begin{equation*}
H_B=\hbar\sum_k\omega_k b_k^{\dag}b_k,
\end{equation*}
where $b_k$ and $b_k^{\dag}$ (annihilation and creation operators corresponding to mode $w_k$, respectively) satisfy the usual Bosonic commutation relation: $[b_k,b_{k'}^{\dag}]=\delta_{k,k'}$. Finally, interaction between system and environment is described by
\begin{equation}
H_I=\hbar\eta\sum_{i=1}^N\sum_k(\sigma_i^-b_k^{\dag}+\sigma_i^+b_k) 
\end{equation}
with $\sigma_i^-=|g\rangle {_i}\langle e|$ (lowering operator at site $i$) and $\sigma_i^+={\sigma_i^-}^{\dag}$ (raising operator at site $i$).
Hence dynamics of $N$-qubit system in interaction picture is governed by the following master equation {\cite{BreuerBook}}:
\begin{eqnarray}\label{master}
\dot{\rho}=\mathcal{L}[\rho]&=& (1+\tau)( 2J^-\rho J^+-\{J^+J^-,\rho\})\cr\cr
&+& \tau(2J^+\rho J^--\{J^-J^+,\rho\})
\end{eqnarray}
where for the sake of simplicity the dissipation rate has been set equal to $1$, 
\begin{equation}
J^-=\sum_{i=1}^N\sigma_i^-,\quad
J^+=\sum_{i=1}^N
\sigma_i^+,
\end{equation}
and
\begin{equation*}
\tau=\frac{1}{e^{\beta \hbar \omega}-1},
\end{equation*}
is the mean photon number of environment with frequency $\omega$. 
\\
\\
Here we are interested in analysing the effect of bath temperature on coherence of the system and also probability of finding a qubit in excited state.   In our analysis, to quantify the coherence of system describing by density matrix $\rho$, we use normalized $l_1$ norm \cite{Baumgratz}
which in basis $\{ |i\rangle  , i=1..d \}$ is defined as follows:
\begin{equation}\label{l1norm}
C(\rho)=||\rho ||_1=\frac{1}{d-1}\sum_{i\neq j} |\langle i|\rho| j\rangle|
\end{equation}
where $i$ and $j$ go from $1$ to $d$. Using this normalized form, the coherence of the maximally coherent state is equal to $1$ \cite{Baumgratz}. For our problem, we fix the basis by product eigenvectors of system Hamiltonian $H_s$ that is $\{ |g\rangle_i, |e\rangle_i \}^{\otimes n}$. To analyse dynamics of coherence in time and effect of bath temperature and system size on that, master equation  \eqref{master} must be solved for system density matrix $\rho$. In next section we explain why this is a cumbersome task (even numerically) and explain our approach for solving the problem. \\
\\
It is also worth noticing that the steady state of the dynamics is not unique. To have a unique steady state, the only operator commuting with all the Lindblad operators must be proportional to identity \cite{Spohn} which is not the case here. Hence master equation \eqref{master} does not confirm a unique steady state. In other words, depending on the initial state, different steady states are expected. Hence in addition to bath temperature and system size, we expect that coherence behaviour depends on the initial state as well. Furthermore, as we will see conservation law which prevents the system to get thermalized, leads to finding the analytical form for steady states.\\
\\
\section{Method} \label{method-sec}
In order to find the coherence of a $N$-qubit system in time, we must solve master equation \eqref{master} for density matrix of the system which belongs to $\mathbb{H}=\mathcal{H}\otimes\mathcal{H^*}$. It is equivalent to solving $\frac{d(d+1)}{2}-1$ set of coupled differential equations for  elements of density matrix $\rho$, where $d=2^N$. Therefore number of coupled differential equations increases exponentially in number of qubits $N$. Here by considering the symmetry and invariant quantities, we show that  number of coupled differential equations becomes polynomial in system size. It is very helpful at least for numerical analysis and also finding the form of steady state.\\
\\
To find the invariant quantities, we first recall that 
the general form of Markovian dynamics generated by $\mathcal{L}$ has the following form in Schrödinger picture: 
\begin{equation}
\dot{\rho}=\mathcal{L}[\rho ]=-i[H,\rho]+\sum_{i}\gamma_i (2L_i\rho L_i^{\dag}-\{L_i^{\dag}L_i,\rho\}),
\end{equation}
with $L_i$ being Lindblad operators. Formal solution of the above master equation for density matrix is given by $\rho(t)=e^{t\mathcal{L}}\rho(0)$. In the same setting dynamics of observable is governed by its adjoint of $\mathcal{L}$ denoted by $\mathcal{L}^{\dagger}$ which is defined as follows: 
\begin{equation}
(A,{\mathcal L}[B])=( {\mathcal L}^{\dagger}[A],B),
\end{equation}
where $A$ and $B$ are any operator on $\mathbb{H}$ and  $(X,Y):=tr(XY)$ is inner product of operators. Using the cyclic properties of trace, form of $\mathcal{L}^{\dag}$ in terms of Lindblad operators $L_i$, and $H$ is given by
\begin{equation}
\mathcal{L}^{\dag}[A]=i[H,A]+\sum_{i}\gamma_i (2L_i^{\dag}A L_i-\{L_i^{\dag} L_i,A\}).
\end{equation}
Furthermore, expectation value of observable $O$ on system must have the same value in Schrodinger and Heisenberg picture:
\begin{equation}
\langle O\rangle=tr(Oe^{t\mathcal{L}}[\rho(0)])=tr(e^{t\mathcal{L^{\dagger}}}[O]\rho(0))=tr(O(t)\rho(0)).
\end{equation}
As the above equality should be valid for any initial state $\rho(0)$, it is concluded that the dynamics of observable $O$ is governed by $\mathcal{L}^{\dag}$:
\begin{equation}
\dot{O}=\mathcal{L}^{\dag}[O].
\end{equation}
Hence any observable that satisfies $\mathcal{L}^{\dag}[O]=0$, is constant of motion and invariant under the dynamics. \\
\\
In the system under consideration, master equation in \eqref{master} is described by two Lindblad operators $J^-$ and $J^+$, Hence: 
\begin{eqnarray} \label{L-adjoint}
\mathcal{L}^{\dag}[O]&=&(1+\tau)(2J^+ O J^--\{J^+J^-,O\})\cr\cr
&+&\tau(2J^- OJ^+ -\{J^- J^+,O\}).
\end{eqnarray}
Using the commutation relation of $J^+$ and $J^-$ with total angular momentum of $N$ spin half particles $J^2$, it is easy to see that $\mathcal{L}^{\dag}[J^2]=0$ and hence $J^2$ is constant of motion. In fact, regarding the Taylor expansion of $e^{t\mathcal L}$ density matrix of the system at arbitrary time is given in terms of successive action of $\mathcal{L}$ on initial state:
\begin{equation}
\rho(t)=\sum_{k}\frac{t^k}{k!}\mathcal{L}^k[\rho(0)].
\end{equation}
Hence if we consider $|j,m\rangle$, the common eigenvector of $J^2$ and 
$J_z$ ( total angular momentum of system in $z$ direction) as initial state, by action of $\mathcal{L}$, $j$ remains constant and Lindblad operators which are lowering and raising operators of $su(2)$ Algebra just change the value of $m$. It is easy to see that 
\begin{eqnarray} \label{L-Pjm}
\mathcal{L}[P_{j,m}]&=&2(1+\tau){c_{j,m}^-}^2(P_{j,m-1}-P_{j,m})\cr\cr
&+&2\tau {c_{j,m}^+}^2(P_{j,m+1}-P_{j,m}),
\end{eqnarray}
where $P_{j,m}:=|j,m\rangle\langle j,m|$ are orthogonal independent operators and $c_{j,m}^{\pm}=(j(j+1)-m(m\pm 1))^{1/2}$. Hence subspace
$\mathbb{H}_{\rho_0}=Span\{P_{j,m}|m=-j,-j+1,\cdots,j-1,j\}$ is the invariant dynamical subspace corresponding to initial state $\rho_0=|j,m\rangle\langle j, m|$. It implies that during the evolution the density matrix of the system remains in a $2j+1$ dimensional subspace. This enables us to reduce the number of differential equations given by master equation \eqref{master} by expanding density matrix of the system in arbitrary time in terms of the basis of dynamical subspace: 
\begin{equation}\label{InitialJM}
\rho(t)=\sum_{m=-j}^j u_m(t) P_{j,m}.
\end{equation}
Above equation and equation \eqref{master} give $2j+1$ differential equations for coefficients $u_m(t)$. For a system of $N$ qubits, the largest value of $j$ is $\frac{N}{2}$, hence for initial state of form $|j,m\rangle$ at most $N+1$ coupled differential equations should be solved which is polynomial in $N$. It makes the numerical analysis much easier and also is helpful in finding the explicit form of the stationary state of the system. \\
\\
So far we have shown that as long as initial state is a common eigenvector of invariant operator $J^2$ and $J_z$ number of coupled differential equation becomes polynomial in system size. This idea can be generalized for other initial states as well. In case initial state is not an eigenstate of $J^2$, we expand it in terms of common eigenvectors of $J^2$ and $J_z$. As an example, let us consider an initial state with single excitation at qubit one: $|\psi (0)\rangle=|e\rangle |g\rangle ^{\otimes N-1}$, which is a typical initial state for analysing excitation transfer. This state is an eigenvector of $J_z$ with eigenvalue $m=-(\frac{N}{2}-1)$ and can be written as a superposition of two states with total angular momentum $\frac{N}{2}$ and $\frac{N}{2}-1$:
\begin{eqnarray}
|\psi(0)\rangle=\frac{1}{\sqrt{N}}&&|j=\frac{N}{2},m=-(\frac{N}{2}-1)\rangle\cr\cr
+\sqrt{\frac{N-1}{N}}&&|j=\frac{N}{2}-1,m=-(\frac{N}{2}-1)\rangle.
\end{eqnarray}
Hence initial density matrix is given by
\begin{eqnarray}\label{rho0single}
\rho(0)&=&\frac{1}{N}(P_{\frac{N}{2},-\frac{N}{2}+1}+(N-1)P_{\frac{N}{2}-1,-\frac{N}{2}+1}\cr\cr
&+&\sqrt{N-1}Q_{\frac{N}{2},m=-\frac{N}{2}+1}),
\end{eqnarray}
in which $P_{j,m}$ is defined after equation \eqref{L-Pjm} and $Q_{j,m}$ is defined as follows:
\begin{equation}
Q_{j,m}=|j,m\rangle\langle j-1,m|+|j-1,m\rangle\langle j,m|,
\end{equation}
where $m$ takes $2j-1$ integer values: $-(j-1)\leq m\leq j-1$. As discussed after equation \eqref{L-adjoint}
dynamics preserves $J^2$ and set of operators $\{P_{j,m}\}$ is an invariant set under $\mathcal{L}$. Similarly, it is easy to see that the set of operators $Q_{j,m}$ with fixed value of $j$ construct an invariant set under the action of $\mathcal{L}$:
\begin{eqnarray}
&&\mathcal{L}[Q_{j,m}]=\cr\cr
&&(1+\tau)(2c_{j,m}^-c_{j-1,m}^-Q_{j,m-1}-({c_{j,m}^-}^2+{c_{j-1,m}^-}^2)Q_{j,m}\cr\cr
&&\tau(2c_{j,m}^+c_{j-1,m}^+Q_{j,m+1}-({c_{j,m}^+}^2+{c_{j-1,m}^+}^2)Q_{j,m}.
\end{eqnarray}
Hence density matrix corresponding to initial state $\rho_0$ in equation \eqref{rho0single}, evolves in a subspace of $\mathbb{H}$ denoted by
$\mathbb{H}_{\rho_0}$ spanned by $3N-1$ independent operators:
$\{P_{j^*,m},P_{j^*-1,m},Q_{j^*,m}\}$, with $j^*=\frac{N}{2}$:

\begin{eqnarray} \label{basis-expansion}
\rho(t)&=&\sum_{m=-j}^j u_m(t)P_{j^*,m}+ \sum_{m=-j+1}^{j-1} v_m(t) P_{j^*-1,m}\cr\cr
&+&\sum_{m=-j+1}^{j-1}w_m (t)Q_{j^*,m}.
\end{eqnarray}
Putting this density matrix in master equation in equation \eqref{master}, we find $3N-1$ differential equations for coefficients in the above equation. Again we find the number of coupled equation to be polynomial in $N$.\\
\\
In Appendix \ref{AppA} we generalized this idea of using invariant quantities to reduce the number of differential equations given by a master equation. We discuss that for a class of Markovian dynamics where Lindblad operators are ladder operators of semi-simple algebra, the Casimir operator is the invariant operator and hence the dynamical subspace is specified by its value. We also discuss how the dimension of dynamical subspace is related to eigenvector of Cartan sub-algebra with highest weight. 
\section{Incoherent Initial State with single excitation}\label{sec:Incoherent}
In this section we consider an initial state with single excitation at qubit $k$ and assume that all the other qubits are in ground state.
From \eqref{l1norm}, it is clear that the initial coherence of the system is zero when the natural basis for computing coherence is 
$\{|e\rangle,|g\rangle\}^{\otimes N}$. We are interested to see if coherence can be generated during the dynamics of the system and effect of environment temperature on that. Furthermore, we want to study the effect of environment temperature on probability of finding excitation on qubits which were initially in ground state.\\
\\
Regarding the arguments in section \ref{method-sec}, the initial state is given by equation \eqref{rho0single} and dynamical subspace is spanned by $3N-1$ independent operators $\{P_{j^*,m},P_{j^*-1,m},Q_{j^*,m}\}$ with $j^*=\frac{N}{2}$. Here for ease of calculation of coherence and also probability of finding  a qubit in excited state, we introduce the following set of operators as basis of dynamical subspace:
\begin{eqnarray}\label{basisDefinition}
&&\Lambda_n=|nE_k\rangle\langle nE_k|\quad 1\leq n\leq N, \cr\cr
&&\Omega_n=|nE_{\not{k}}\rangle\langle nE_{\not{k}}|\quad 0\leq n\leq N-1, \cr\cr
&&\chi_n=|nE_k\rangle\langle nE_{\not{k}}|+|nE_{\not{k}}\rangle\langle nE_k|\quad 1\leq n\leq N-1.\cr\cr
&&
\end{eqnarray}
By $|nE_k\rangle$ we represent a normalized state which has $n$ number of excitation when one of them is in qubit $k$. Hence in this notation the initial state is represented by $|1E_k\rangle$. $|nE_{\not{k}}\rangle$, represents a normalized state with $n$ number of excitation in which qubit $k$ is not excited. by this notation, $\Omega_0=|G\rangle\langle G|=|g\rangle\langle g|^{\otimes N}$ represent the state where all the qubits are in ground state. Operators in  \eqref{basisDefinition} form $3N-1$ independant operators and subspace $\mathbb{H}_{|1E_k\rangle\langle 1E_k|}=Span\{\Lambda_n,\Omega_n,\chi_n\}$ is invariant subspace under $\mathcal{L}$. Therefore the density matrix at arbitrary time corresponding to initial state $|1E_k\rangle$ is given by
\begin{equation}\label{rhoIncoherent}
\rho(t)=\sum_{n=1}^N a_n (t) \Lambda_n+\sum_{n=0}^{N-1}b_{n}(t)\Omega_n+\sum_{n=1}^{N-1}c_n(t)\chi_n.
\end{equation}
Regarding the definition in equation \eqref{l1norm}, coherence of this state is given by
\begin{equation}\label{Cincoherent}
C(\rho(t))=\frac{\sum_{n=1}^N\binom{N-1}{n-1}(|a_n(t)|+|b_{n-1}(t)|)+\frac{2}{n}\sqrt{f_n}|c_n(t)|)-1}{2^N -1}.
\end{equation}
Using equation (\ref{rhoIncoherent}) it is straightforward to show that the probability of finding qubit $l\neq k$ (which was initially in ground state) in excited state is given by:
\begin{equation}\label{pl}
p_l(t)=\frac{\sum_{n=1}^Nn(a_n(t)+b_{n-1}(t))-1}{N-1},  \quad l\neq k.
\end{equation} 
To find the explicit behaviour of coherence and probability of having an excitation at each qubit, coefficients
$a_n(t)$, $b_n(t)$ and $c_n(t)$ must be found. In order to find the differential equations governing the dynamics of these coefficients, we first note that 
\begin{eqnarray}\label{CS-Incoherent}
\mathcal{L}[\Lambda_n]&=&(1+\tau)\{f_{n-1}\Lambda_{n-1}-(f_n-N+2n)\Lambda_n\cr\cr
&+&\Omega_{n-1}+\sqrt{f_{n-1}}\chi_{n-1}-\frac{1}{2}\sqrt{f_n}\chi_n\}\cr\cr
&+&\tau\{f_n(-\Lambda_n+\Lambda_{n+1})-\frac{1}{2}\sqrt{f_n}\chi_n\},\cr\cr
\mathcal{L}[\Omega_n]&=&(1+\tau)\{f_n(\Omega_{n-1}-\Omega_n)-\frac{1}{2}\sqrt{f_n}\chi_n\}\cr\cr
&+&\tau\{\Lambda_{n+1}-(f_{n+1}+1)\Omega_n+f_{n+1}\Omega_{n+1}+\cr\cr
&-&\frac{1}{2}\sqrt{f_n}\chi_n+\sqrt{f_{n+1}}\chi_{n+1}\},\cr\cr
\mathcal{L}[\chi_n]&=&(1+\tau)\{\sqrt{f_n}(-\Lambda_n+2\Omega_{n-1}-\Omega_n)\cr\cr
&+&\sqrt{f_{n-1}f_n}\chi_{n-1}-\frac{2f_n-N+2n}{2}\chi_n\}\cr\cr
&+&\tau\{\sqrt{f_n}(2\Lambda_{n+1}-\Lambda_n-\Omega_n)-\frac{2f_n+N-2n}{2}\chi_n\cr\cr
&+&\sqrt{f_nf_{n+1}}\chi_{n+1}\},
\end{eqnarray}
with $f_n:=n(N-n)$. Using equations \eqref{master}, \eqref{rhoIncoherent}, \eqref{CS-Incoherent} and  the facts that operators $\Lambda_n$, $\Omega_n$ and $\chi_n$ are independent, a set of $3N-1$ first order coupled differential equations for coefficients $a_n(t)$, $b_n(t)$ and $c_n(t)$ are found. We show these set of equations in the following compact form:
 \begin{equation} \label{incoherent-M-def}
\frac{d}{dt}|v(t)\rangle=\mathcal{M}|v(t)\rangle,
\end{equation}
\begin{figure}[t]
\centering
       \vspace{-0.2 cm}
   \includegraphics[scale=0.4]{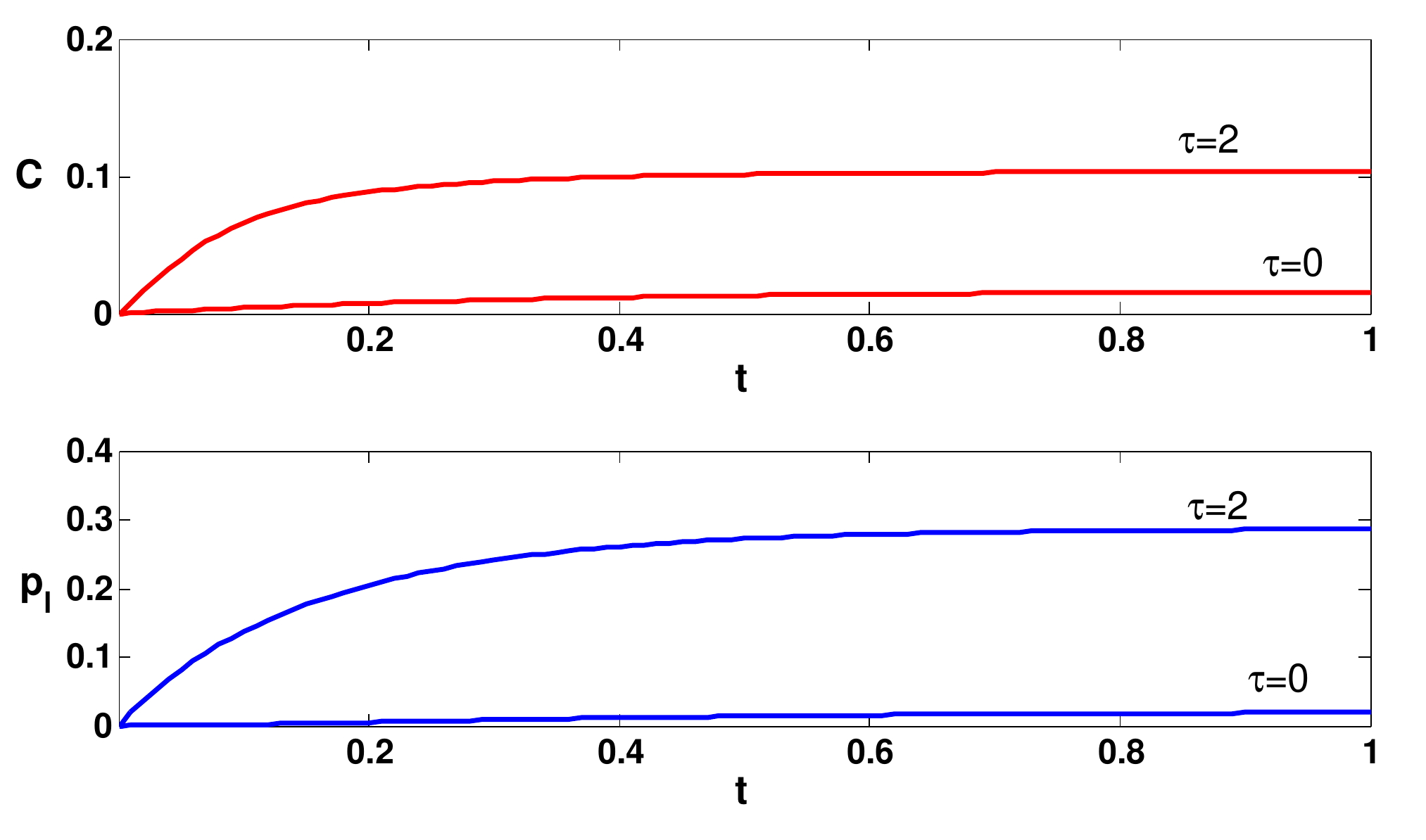}
       \vspace{-0.5 cm}
 \centering
\caption{Top: Coherence versus time at $\tau=0$ and $\tau=2$. Bottom: Probability of having an excitation at qubit $l\neq k$ which is initially in ground state versus time for $\tau=0$ and $\tau=2$. In both figures, initial state is incoherent with single excitation at qubit $k$ and $N=7$.} 
\label{Incoherent_Coherence_Probability_vs_time}
\end{figure}
where $|v(t)\rangle$ is the vector of coefficients 
\begin{equation} \label{incoherent-coef-vec}
|v(t)\rangle=(a_1(t),\cdots a_N(t),b_0(t),\cdots b_{N-1}(t),c_1(t)\cdots c_{N-1}(t))^T
\end{equation}
and the elements of non-positive matrix $\mathcal{M}$ is given by the set of equation in \eqref{CS-Incoherent} (See Appendix \ref{AppB} for more details). Hence coefficients at any arbitrary time are given by
\begin{equation}
|v(t)\rangle=e^{t\mathcal{M}}|v(0)\rangle.
\end{equation}
This equation is used to find $|v(t)\rangle$ numerically. Having $|v(t)\rangle$ or equivalently coefficients in equation \eqref{rhoIncoherent} we find coherence and probability of having excitation at one qubit (equations \eqref{Cincoherent} and \eqref{pl}, respectively) in terms of time. Top plot in figure (\ref{Incoherent_Coherence_Probability_vs_time}), shows behaviour of coherence versus time. Initial state is incoherent state with excitation at qubit $k$. During the evolution initial excitation at qubit $k$ may be lost by emitting a photon to the environment and then another qubit get excited by absorbing this energy from the environment which is excitation transfer through the environment. If environment is at non-zero temperature there is also the possibility that a qubit get excited by absorbing one of the photons of environment. Indeed as time passes the chance that a qubit get excited increases. This results in appearance of more terms in form of superposition or more precisely coherence in the system. Hence increase of coherence in time is in accordance with our intuition. Bottom plot of figure (\ref{Incoherent_Coherence_Probability_vs_time}) also confirms that the probability of finding qubit $l\neq k$ in excited state increases in time. While this figure is plotted for $N=7$ and two fixed value of environment mean photon number, namely $\tau=0$ and $\tau=2$, similar behaviour is seen for other system sizes and temperatures as well. \\
\\
\begin{figure}[t]
\centering
       \vspace{-0.5 cm}
   \includegraphics[scale=0.4]{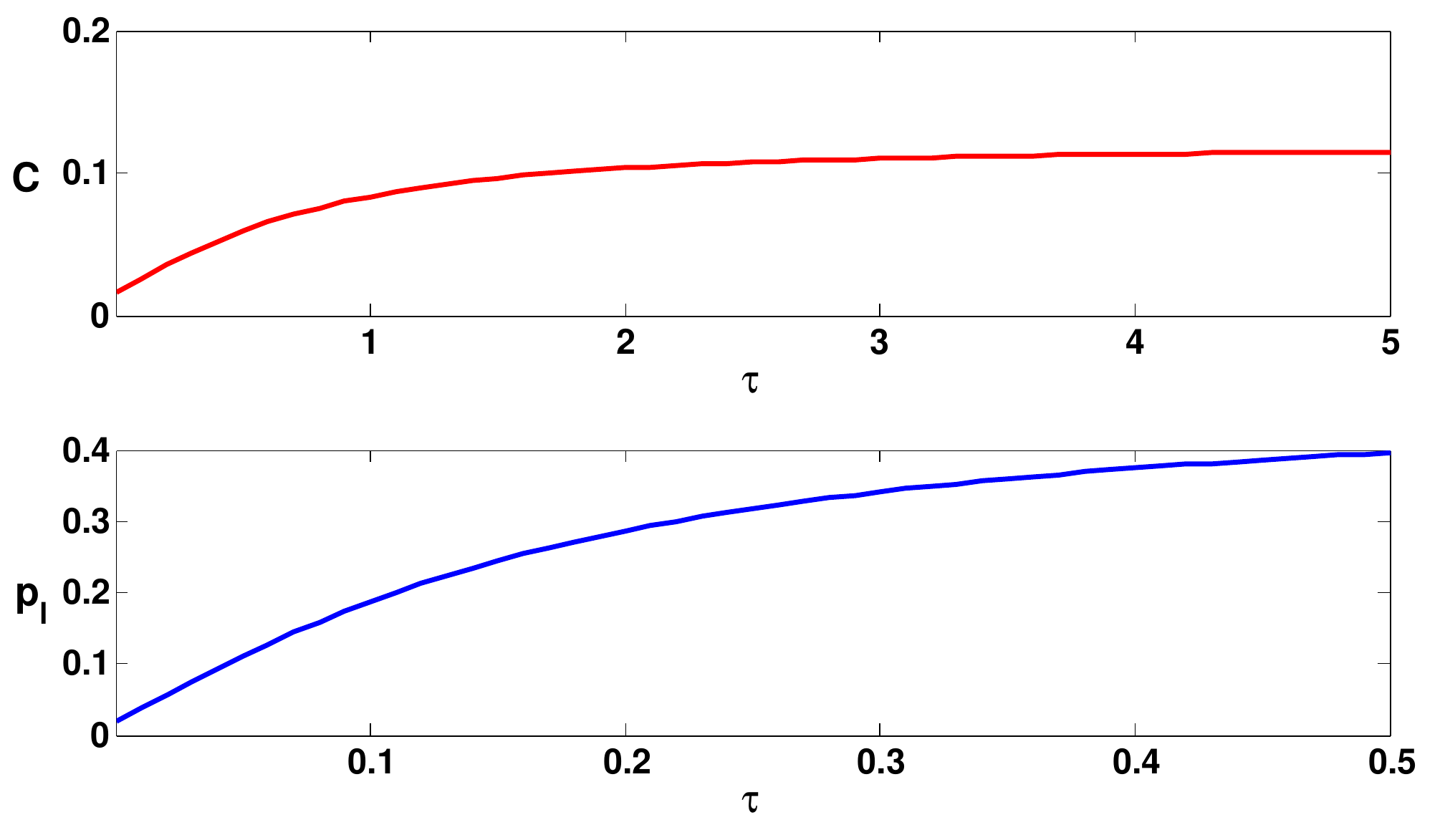}
       \vspace{-0.5 cm}
 \centering
\caption{ Top: Coherence versus bath mean photons number $\tau$. Bottom: Probability of having qubit $l\neq k$ in excited state which is initially in ground state versus bath mean photon number $\tau$. In both figures initial state is incoherent state with single excitation at qubit $k$, time is fixed at $t=1.8$ and $N=7$.} 
\label{Incoherent_Coherence_Probability_vs_tau}
\end{figure}
Interestingly, bath temperature has positive role in generating coherence in the system. In figure (\ref{Incoherent_Coherence_Probability_vs_tau}) for a system of $N=7$ qubits, behaviour of coherence (top figure) and probability of having excitation at qubits which were initially in ground state (bottom figure), are shown at some instant time $t=1.8$ versus bath mean photon number $\tau$. As it is seen coherence increases by increase of bath mean photon number $\tau$ to its saturated value. Actually, as bath temperature increases, the mean photon number related to mode $\omega$ increases. Since this frequency corresponds to the difference in system energy levels, by increasing the bath temperature, the chance that a qubit absorbs energy and get excited increases. Hence probability of finding a qubit in excited state and also coherence increases versus temperature. \\
\\
As it is seen in figure (\ref{Incoherent_Coherence_Probability_vs_time}) coherence and probability of finding a qubit in excited state increase in time and then saturate to a finite value. In fact, since symmetry imposes a constraint on the system to remain in a subspace of the whole Hilbert space, there is an upper bound for the ultimate value of coherence and probability of finding a qubit in excited state which can be found by analysing the steady state of the system. 
The stationary behaviour of the system, is governed by zero-eigenvectors of matrix $\mathcal{M}$. Vector $|v(t)\rangle$ is expressed in terms of  
the eigenvectors of matrix $\mathcal{M}$ as follows:
\begin{equation}
|v(t)\rangle=\sum_{i=1}^{3N-1}e^{t\lambda_i}|r_i ^{(\lambda_i)}\rangle\langle l_i  ^{(\lambda_i)}|v(0)\rangle,
\end{equation}
in which $\lambda_i$s are the eigenvalues of $\mathcal{M}$ and $|r_i^{(\lambda_i)}\rangle$s ($\langle l_i^{(\lambda_i)}|$s) are right (left) eigenvectors of $\mathcal{M}$ with eigenvalue $\lambda_i$. Since matrix $\mathcal{M}$ is non-positive, the stationary state of the system is described
in terms of right and left eigenvectors of matrix $\mathcal{M}$ with zero eigenvalues:
\begin{equation}
|v(\infty)\rangle=|r_1^{(0)}\rangle\langle l_1^{(0)}|v(0)\rangle+|r_2^{(0)}\rangle\langle l_2^{(0)}|v(0)\rangle.
\end{equation}
Density matrices corresponding to two right eigenvectors of $\mathcal{M}$ with zero eigenvalue are given by (For explicit form of right and left eigenvectors of $\mathcal{M}$ see Appendix \ref{AppB}): 
\begin{eqnarray}\label{incoh-steady-rho}
&&\rho_{1}=\frac{1-\nu}{(1+\nu)(1-\nu^N)}\left(\sum_{n=1}^N\nu^n\Lambda_n+\sum_{n=0}^{N-1}\nu^n\Omega_n\right)\cr\cr
&&\rho_{2}=\frac{1-\nu}{\nu(1-\nu^{N-1})}\sum_{n=1}^ {N-1} \nu^n|\psi_n\rangle\langle\psi_n|,
\end{eqnarray}
with
\begin{equation}
\nu:=\frac{\tau}{1+\tau}=e^{-\beta\hbar\omega},
\end{equation}
and
\begin{equation}
|\psi_n\rangle=\frac{1}{\sqrt{N}}(\sqrt{N-n}|nE_k \rangle-\sqrt{n}| |nE_{\not{k}}\rangle).
\end{equation}
Using left zero eigenvectors and knowing that the only non-zero element of $|v(0)\rangle$ is its first element, we find that the steady state of the system which corresponds to $|v(\infty)\rangle$ is given by
\begin{equation}\label{I-S-rho}
\rho(\infty)=\alpha\rho_{1}+(1-\alpha)\rho_{2},
\end{equation}
where
\begin{equation}
\alpha:=\langle l_1^{(0)}|v(0)\rangle=\frac{(1+\nu)(1-\nu^{N})}{N(1-\nu^{N+1})}.
\end{equation}
By having the explicit form of system density matrix in steady state, we can analyse how quantum features of system in steady state, like coherence, behave in terms of bath temperature and how they scale with system size. As is shown in figure (\ref{Incoherent_Coherence_Probability_vs_time}), coherence in the system increases in time and saturates as system gets to its steady state. The saturated value of coherence dependants on the bath temperature (or equivalently environment mean photon number) and also system size. By using equations \eqref{Cincoherent} and \eqref{I-S-rho}, we find coherence in the steady state which is the maximum value of coherence generated in the system:
\begin{eqnarray}
C(\rho(\infty))&=&\frac{1}{2^N-1}( \alpha \frac{(1-\nu)(1+\nu)^{N-1}}{(1-\nu^{N})}\cr\cr
&+&4 (1-\alpha)  \frac{N-1}{N} \frac{(1-\nu)(1+\nu)^{N-2}}{(1-\nu^{N-1})}-1 ) .\cr\cr
&&
\end{eqnarray}
\begin{figure}[t]
\centering
       \vspace{-0.2 cm}
   \includegraphics[scale=0.4]{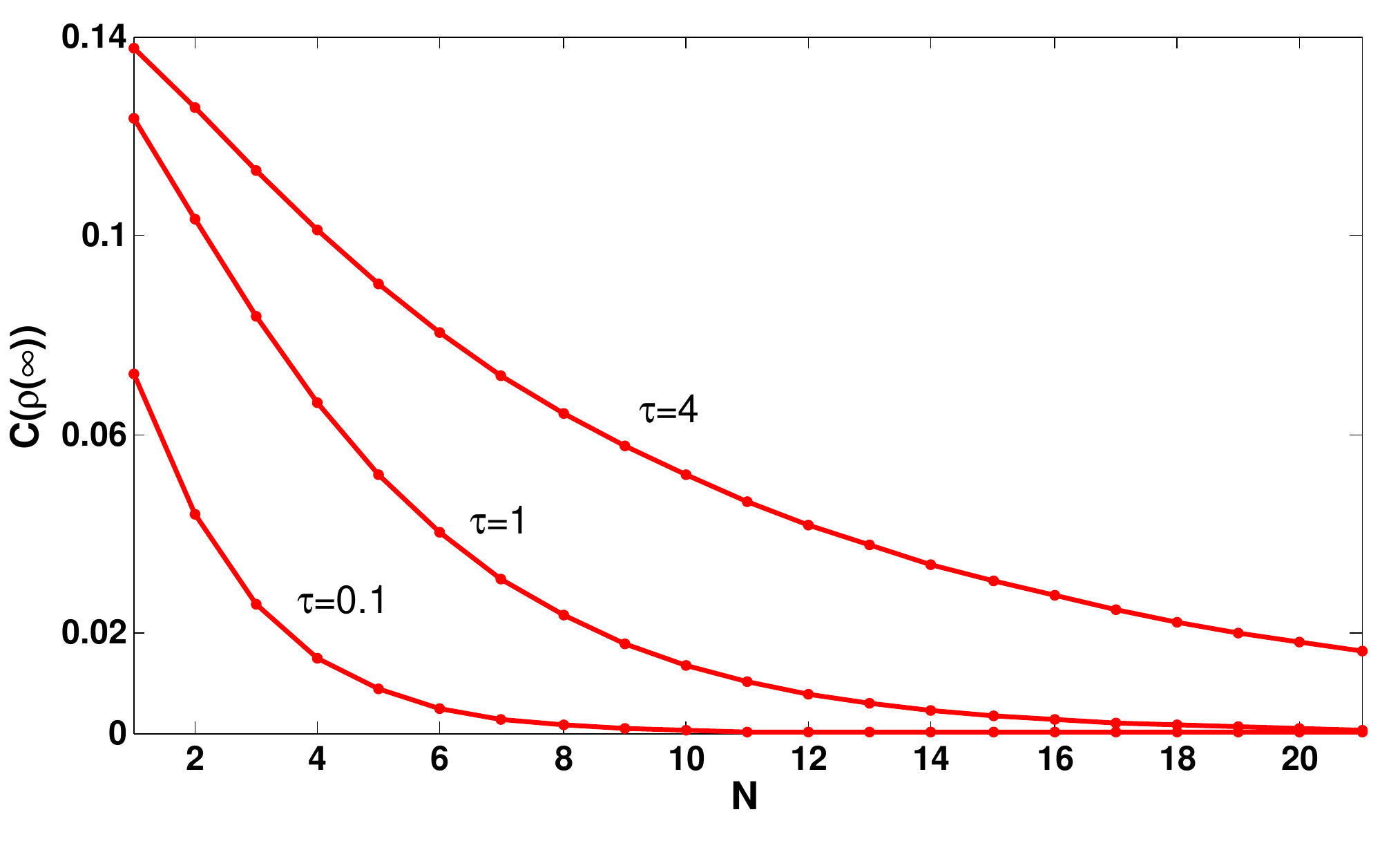}
       \vspace{-0.5 cm}
 \centering
\caption{Coherence of steady state versus system size $N$ for $\tau=0.1$, $\tau=1$ and $\tau=4$ from bottom to top. Initial state is incoherent state with single excitation at qubit $k$ and $N=7$.} 
\label{Incoherent_steady_vs_size}
\end{figure}
In figure (\ref{Incoherent_steady_vs_size}) it is shown that the maximum value of coherence generated in the system, decreases by system size and eventually approaches zero. Hence in such systems no Macroscopic coherence will be observed. It is also expected that for lower values of bath temperature or equivalently smaller values of $\tau$, coherence decays faster as system size increases. Scaling of coherence with system size in two limits of low temperature ($T\rightarrow 0$, or equivalently $\nu\rightarrow 0$) and high temperature ($T\rightarrow\infty$, or equivalently $\nu\rightarrow 1$) has the following form:
\begin{eqnarray}
\lim_{\nu\rightarrow 0}C(\infty)&=&  \frac{1}{2^N-1} \frac{(N-1)(3N-4)}{N^2},\cr\cr
\lim_{\nu\rightarrow 1}C(\infty)&=&\frac{2^N}{ { \left(2^N-1 \right) } \left(N+1\right)}.
\end{eqnarray}
It is also interesting to look at the probability of finding a qubit in excited state in these two limits: 
\begin{eqnarray}
\lim_{\nu\rightarrow 0}p_{l\neq k}(\infty)&=&\frac{1}{N^2},\cr\cr
\lim_{\nu\rightarrow 1}p_{l\neq k}(\infty)&=&\frac{1}{2}.
\end{eqnarray}
When system gets to its steady state, in the limit of low bath temperature, probability of finding a qubit in excited state, decreases by system size as $\frac{1}{N^2}$. That is because in this limit number of environment photons are not large enough that each qubit has a chance of absorbing energy and get excited. But by increasing the temperature, all the qubits have possibility to absorb energy from environment and become excited with probability $\frac{1}{2}$, no matter what the system size is. 

\section{Coherent Initial State with single excitation}\label{sec:Coherent}
In previous section, we showed that if system of $N$ qubits is initially in incoherent state with single excitation, during the interaction with environment at temperature $T$, coherence in the system and also the probability of finding a qubit (initially in ground state) in excited state, increases. Getting motivated by this result, we ask whether or not by preparing the initial state in coherent state, we can achieve higher value of coherence in steady state and also increase in probability of finding a qubit in excited state. Hence as initial state, we consider a uniform superposition of states with single excitation. We represent a normalized uniform superposition of pure states with $n$ excitation by $|nE\rangle$. When all the qubits are in ground state $n=0$ hence such a state is represented by $|G\rangle$. Using this notation the initial coherent state with single excitation is represented by $|1E\rangle$ and its coherence is given by $\frac{N-1}{2^N-1}$. This state is eigenstate of $J^2$ and $J_z$ which regarding the notation used in section \ref{method-sec} is represented by $|j=\frac{N}{2}, m=1-\frac{N}{2}\rangle$. 
Hence dynamical subspace is spanned by $P_{j=\frac{N}{2},m}$ ($m$ is an integer between $\pm j$) and its dimension is $N+1$. We denote this subspace by $\mathbb{H}_{|1E\rangle\langle 1E|}$. For the aim of calculating coherence in basis $\{|g\rangle,|e\rangle\}^{\otimes N}$ and also finding the probability of having excited qubit, it is more convenient to use a new notation for the basis of dynamical subspace: $\Gamma_n:=|nE\rangle\langle nE|=P_{j=\frac{N}{2},m=n-\frac{N}{2}}$. Hence equation \eqref{L-Pjm} rewritten in terms of $\Gamma_n$ becomes:
\begin{eqnarray}\label{LGamma}
\mathcal{L}[\Gamma_n]&=&2(1+\tau)(f_n+n)\left(\Gamma_{n-1}-\Gamma_n\right)\cr\cr
&+&2\tau(f_{n+1}+n+1)\left(\Gamma_{n+1}-\Gamma_n\right).
\end{eqnarray}
\\
And the dynamical subspaces is given by:
\begin{equation}
\mathbb{H}_{|1E\rangle\langle 1E|}=\mathit{Span}\{\Gamma_n, n=0,\cdots, N\}
\end{equation}
Therefore, system at arbitrary time $t$ is described as follows:
\begin{equation}\label{coherent-coef-definition}
\rho(t)=\sum_{n=0}^Nd_n(t)\Gamma_n,
\end{equation}
and coherence of the system is given by
\begin{equation}\label{CCoherent}
C(\rho(t))=\frac{\sum_{n=0}^N\binom{N}{n}d_n(t)-1}{2^N-1}.
\end{equation}
While initially each qubit is in excited state with probability $\frac{1}{N}$, the probability of finding a qubit excited at time $t$ is given by:
\begin{equation}
p_l(t)=\frac{1}{N}\sum_{n=0}^Nnd_n(t).
\end{equation}
To find the coefficients $d_n(t)$, we use the same technique of previous section. Using equation \eqref{coherent-coef-definition} and master equation in \eqref{master} a set of $N+1$ differential equations is found for coefficients $d_n(t)$ which is summarized as follows:
\begin{equation}
|\dot{u}(t)\rangle=\mathcal{M}'|u(t)\rangle,
\end{equation}
where $|u(t)\rangle$ is vector of coefficients $d_n(t)$:
 \begin{equation}\label{coherent-coef-vec}
|u(t)\rangle:=(d_0(t),d_1(t),\cdots,d_N(t))^T.
\end{equation}
Elements of the non-positive matrix $\mathcal{M'}$ are given by equation \eqref{LGamma} as follows:
\begin{eqnarray} \label{Mprime-Elements}
&&\mathcal{M'}_{n,n}=-2(1+\tau)(f_n+n)-2\tau(f_{n+1}+n+1),\cr\cr
&&\mathcal{M'}_{n+1,n}=2\tau(f_{n+1}+n+1),\cr\cr
&&\mathcal{M'}_{n-1,n}=2(1+\tau)(f_n+n).
\end{eqnarray} 
Thus solution of the set of coupled differential equations is described as follows:
\begin{equation}
|u(t)\rangle=e^{\mathcal{M'}t}|u(0)\rangle.
\end{equation}
\begin{figure}[t]
\centering
       \vspace{-0.5 cm}
   \includegraphics[scale=0.4]{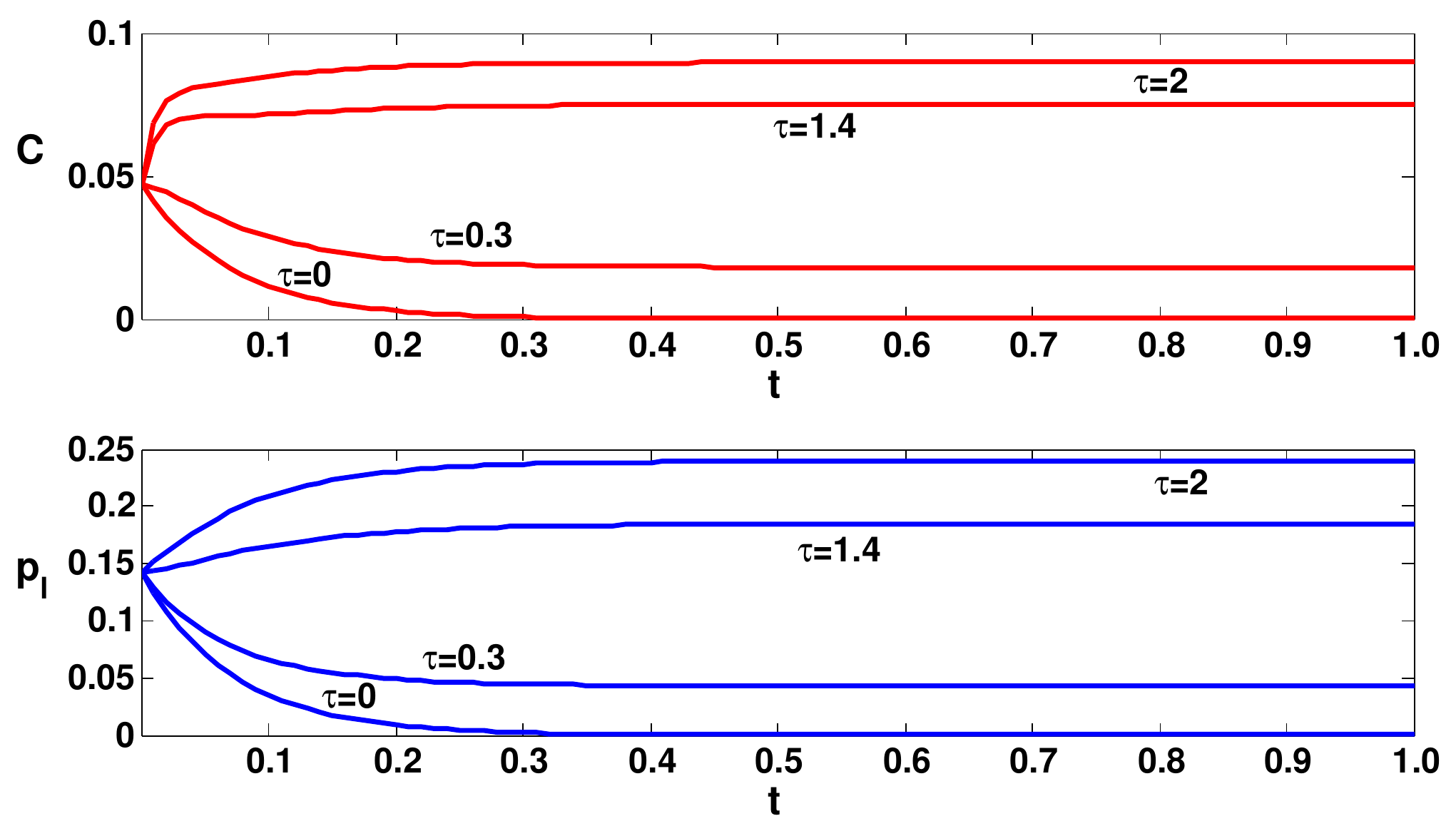}
       \vspace{-0.5 cm}
 \centering
\caption{ Top: Coherence versus time. Bottom: Probability of finding an arbitrary qubit in excited state versus time. In both figures initial state is coherent state, $N=7$ and $\tau=0,0.3,1.4$ and $2$ from bottom to top } 
\label{Coherent_Coherence_Probability}
\end{figure}
Figure (\ref{Coherent_Coherence_Probability}) shows the behaviour of coherence and probability of having a qubit in excited state versus time for various values of $\tau$ and $N=7$. As it is seen in this figure depending on the bath mean photon number (or equivalently bath temperature) the steady value of coherence is either larger or smaller than initial coherence. In fact there is a critical value of $\tau$ denoted by $\tau_c$ that when $\tau<\tau_c$, coherence of steady state is less than the initial coherence and when $\tau>\tau_c$, system is more coherent in its steady state compared to its initial state. This can be explained by considering the fact that quantum features of the system are result of two phenomena: dissipating energy to the environment and absorbing energy from environment. It is always probable that system loses its excitation by emitting energy to the environment and tends to become in ground state with zero coherence. In the meanwhile each qubit can get excited by absorbing energy from the environment which results in increase of coherence. When bath mean photon number is not large enough that absorbing energy compensates for dissipation, coherence and probability of having a qubit in excited state decreases in time. But when mean photon number of environment is large enough that energy absorption becomes dominant effect, coherence and also probability of finding a qubit in excited state increases during the evolution. It is worth noticing that when initial state has zero coherence (as discussed in previous section) even in zero temperature coherence increases in time as that initial single excitation can hop to other qubits through the environment which causes increase of coherence. \\
\\
To estimate the value $\tau_c$, it is required to find the steady state of the system. Representing left and right eigenvectors of $\mathcal{M'}$ respectively by $|r'^{(\lambda_i)}_i\rangle$ and $\langle l'^{(\lambda_i)}_i|$, vector of coefficients $|u(t)\rangle$ at arbitrary time $t$ is given by
\begin{equation}
|u(t)\rangle=\sum_{i=1}^{N+1}e^{t\lambda_i}|r'^{(\lambda_i)}_i\rangle\langle l'^{(\lambda_i)}_i|u(0)\rangle.
\end{equation}
Matrix $\mathcal{M'}$ is a non-positive matrix with one zero eigenvalue. Steady state of the system is described in terms of right and left eigenvectors of $\mathcal{M'}$ with zero eigenvalue (See appendix \ref{AppC} for details). That is,
\begin{equation}
\rho(\infty)=\frac{1-\nu}{1-\nu^{N+1}}\sum_{n=0}^N\nu^n|nE\rangle\langle nE|,
\end{equation}
with
$\nu=\frac{\tau}{1+\tau}=e^{-\beta\hbar\omega}$. Hence by using \eqref{CCoherent} we find coherence in the steady state to be:
\begin{equation}\label{CoherentCinfinity}
C(\rho(\infty))=\frac{1}{2^N-1}\left(\frac{(1-\nu)(1+\nu)^N}{1-\nu^{N+1}}-1\right).
\end{equation}
\begin{figure}[t]
\centering
       \vspace{-0.5 cm}
   \includegraphics[scale=0.4]{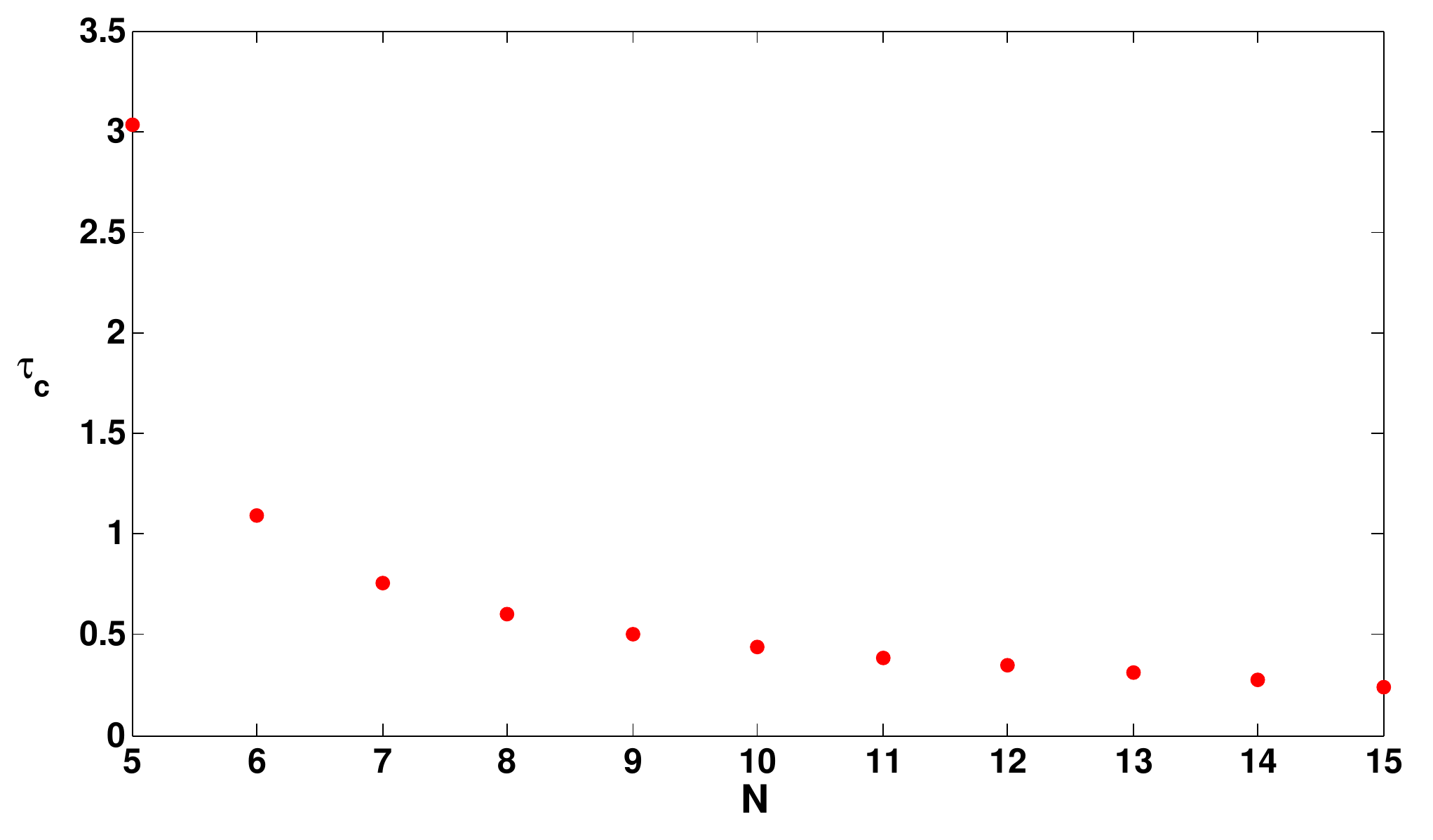}
       \vspace{-0.5 cm}
 \centering
\caption{Critical value of bath mean photon number  versus system size $N$ when initial state is coherent.} 
\label{Coherent_criticalTemp}
\end{figure}
Using the above equation we numerically estimate the value of $\tau_c$, by finding bath mean photon number beyond which coherence in steady state is larger than initial coherence. Figure  (\ref{Coherent_criticalTemp}) shows the behaviour of $\tau_c$ versus system size $N$. As decrease of coherence in terms of system size is much faster for initial state compared to steady state, when system size increases, smaller value of bath mean photon number is sufficient for resulting coherence larger than initial coherence. Hence as it is seen in figure (\ref{Coherent_criticalTemp}), $\tau_c$ is decreasing function of system size.\\
\\
In steady state, probability of finding a qubit in excited state is given by:
\begin{equation}
p(\infty)=\frac{\nu}{N}\frac{1-(N+1)\nu^N+N\nu^{N+1}}{(1-\nu^{N+1})(1-\nu)},
\end{equation}
which shows its dependence on system size $N$ and mean photon number of environment $\tau$. In the limit of zero bath temperature, $\tau\rightarrow 0$ or $\nu\rightarrow 0$, we have $lim_{\nu\rightarrow 0}p_l(\infty)=0$. In this limit system just loses its excitation to the environment and has no chance to absorb energy from environment. In the limit of hight bath temperature, $\tau\rightarrow\infty$ and thus $\nu\rightarrow 1$. Simple calculation show that in this limit $lim_{\nu\rightarrow 1 }p_l(\infty)=\frac{1}{2}$ which is independent of $N$. In this limit no matter how large the system is, all the qubits have the possibility to get excited by absorbing a photon from the environment. 
\section{Conclusion}\label{sec:Conclusion}
In this work we have considered a system of qubits interacting with each other through a common bath at non-zero temperature. With the aim of studying the dynamics of quantum features of the system such as coherence, we have introduced a technique to simplify solving the master equation for a system with arbitrary number of particles. 
We have discussed that in principle number of coupled differential equations that must be solved to find system density matrix, increases exponentially with system size. With the technique introduced in this work, number of coupled differential equations to be solved becomes polynomial in system size. This has been done by considering the system symmetry and corresponding constant of the motion which in this set up is total angular momentum. In fact, by recognizing the invariant quantities we determine the subspace of Hilbert space which the initial state goes through during the evolution. By working in this subspace number of parameter characterizing the state and hence number of differential equations decreases. In addition to that, having evolution in a specific subspace of Hilbert space, clarifies why the steady state of the system is not a thermal state. It is worth noticing that this method is applicable for reducing the complexity of solving large class of master equations. A generalization of this technique for a class of dynamics where Lindblad operators are ladder operators of a semi-simple Algebra is discussed in appendix \ref{AppA}. In this class of dynamics, constants of the motion are Casimir operators of the semi-simple Algerba.\\ 
\\
We have used  the mentioned technique to study the dynamics of coherence. We have shown that two factors play important role in determining the behaviour of coherence in time and also in steady state: initial state and bath temperature. We have also shown that these two factors affect the dynamics of probability of finding a qubit in excited state. In particular, we have shown that when initial state is incoherent state with single excitation at one of the qubits, coherence and probability that a qubit (initially in ground state) gets excited, both increase in time. We have also studied the case that initial state has single excitation distributed uniformly among all qubits and hence has non zero coherence. We have demonstrated that for this initial state, increase or decrease of coherence in time, depends on the bath temperature. If the bath temperature is high enough that rate of absorbing energy from the environment overcomes the dissipation rate, coherence increases in time. Otherwise system gradually loses its initial coherence to a lower value. Hence when initial state is coherent, there exist a critical value of bath temperature that one should go beyond that in order to prevent the loss of coherence and induce more coherence in the system. Interestingly, we see the same behaviour for the probability of finding a qubit in excited state. Our results concerning the similarity between the behaviour of probability of having excitation at each qubit and coherence confirms that by increasing (decreasing) coherence, the probability of finding a qubit in excited state increases (decreases). \\
\\
\begin{figure}
\subfloat{\includegraphics[scale=0.4]{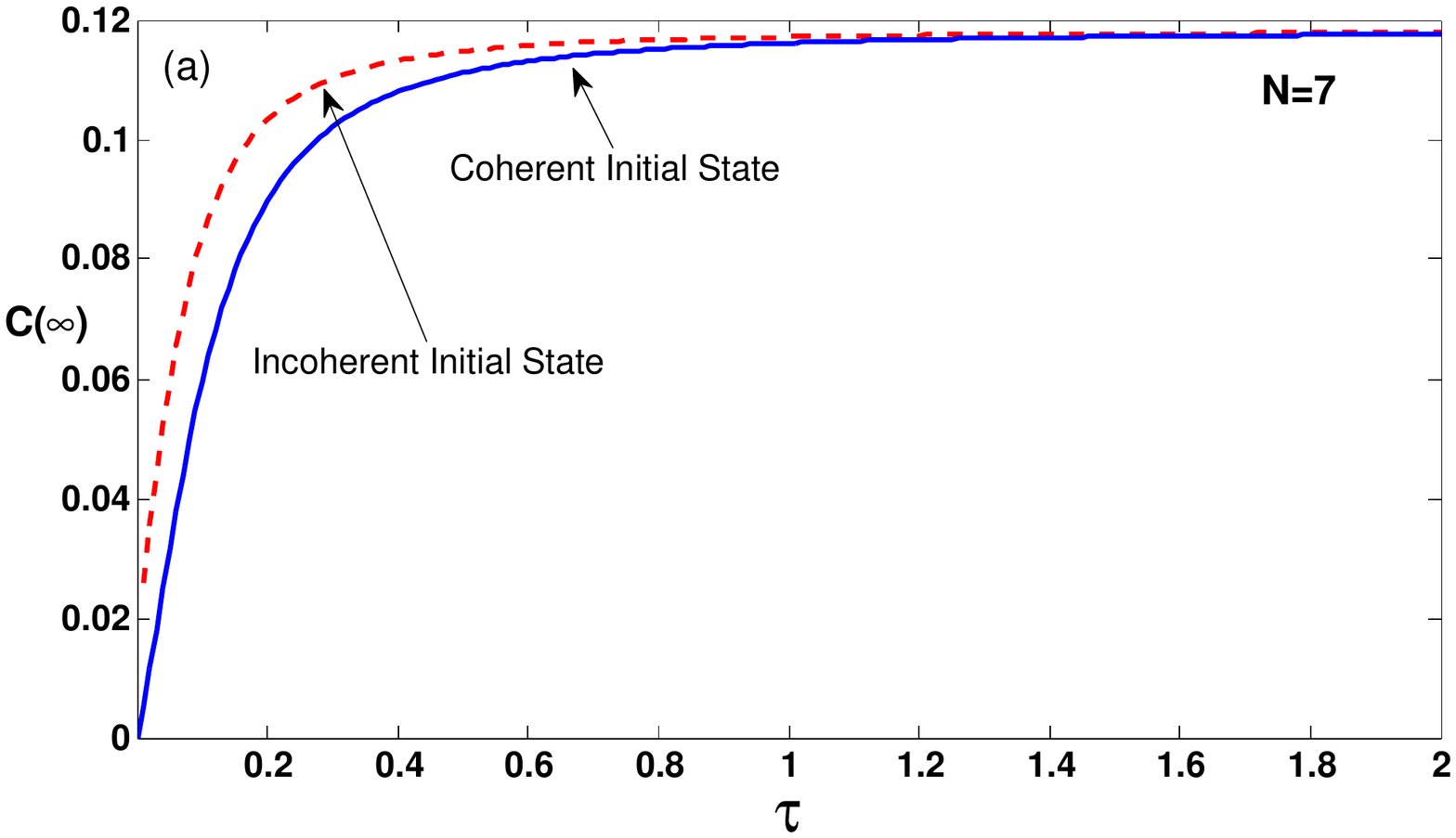}}\\
\subfloat{\includegraphics[scale=0.4]{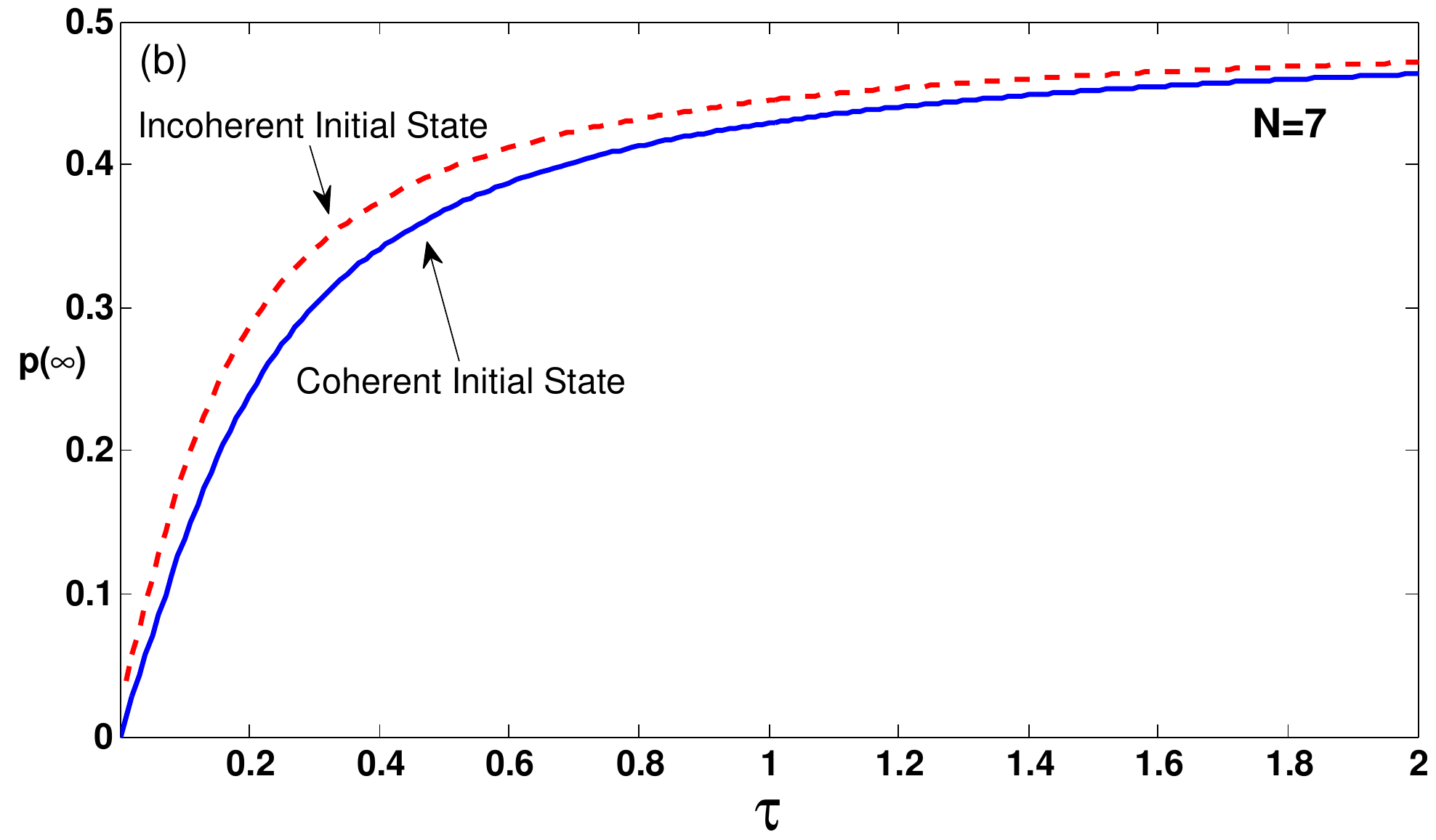}}
\caption{(a): Coherence in steady state versus bath mean photon number for incoherent initial state (dashed red line) and coherent initial state (solid blue line). (b): Probability of finding a qubit in excited state when system is in steady state versus bath mean photon number for incoherent initial state (dashed red line) and coherent initial state (solid blue line). System size is $N=7$. }
\label{SteadyCoherenceProbability}
\end{figure}
These results, apart from being useful to gain deeper insight about coherence, provide helpful information for initializing set ups to achieve specific aims. For example if high value of coherence in steady state is required which initial state should be chosen? What if we want to have  maximum probability of finding a qubit in excited state? How the temperature of bath should be set? In fact by comparing coherence of steady state for different choices of initial state in terms of $\tau$ we can answer these questions. As shown in figure (\ref{SteadyCoherenceProbability}) (top plot) for small values of bath temperature (corresponding to small values of $\tau$), incoherence initial states achieve higher value of coherence in steady state. As temperature increases, both initial conditions lead to the same value of coherence in steady state. Same happens for probability of finding a qubit in excited state when system is in steady state (figure (\ref{SteadyCoherenceProbability}) bottom plot). Though naively it may be expected that interaction with thermal bath destroys the coherence, figure (\ref{SteadyCoherenceProbability}) also suggests that temperature has positive role in inducing coherence in the system and increasing the chance of distributing excitation in the system. 
\acknowledgments
L. M. acknowledges financial support by Sharif University of
Technology’s Office of Vice President for Research and hospitality
of the Abdus Salam International Centre for Theoretical
Physics (ICTP) where parts of this work were completed. This work was partially supported by Sharif University of Technology's Office of Vice President for Research under Grant No. G930209.
\appendix
\section{Markovian dynamics generated by ladder operators of a semi-simple Algebra}\label{AppA}
First we remind that, for any semi-simple algebra $\mathcal{A}$, Cartan subalgera is defined as the maximal Abelian subalgbera of $\mathcal{A}$. That is, basis of Cartan subalbegra denoted by $H_i$s commute: 
\begin{equation*}
[H_i,H_j]=0\quad i,j=1\cdots l,
\end{equation*}
where $l$ defines the rank of algerba $\mathcal{A}$. Hence all the $H_i$s are simultaneously diagonalizable. Actually, for each $H_i$, other basis of Cartan subalgebra, namely $H_{j\neq i}$, are eigenvectors with zero eigenvalue. To make a basis for algebra $\mathcal{A}$, other eigenvectors are required which are denoted by $E_{\alpha}$:
\begin{equation}\label{HE}
[H_i,E_{\alpha}]=\alpha_{(i)}E_{\alpha},
\end{equation}
where at least one of the $\alpha_{(i)}$ is non vanishing and root $|\alpha\rangle$ is defined as a vector with components $\alpha_{(i)}$. It is known that if $|\alpha\rangle$ is a root, $|-\alpha\rangle$ is a root as well. That is, $[H_i,E_{-\alpha}]=-\alpha_{(i)}E_{-\alpha}$. Finding a representation for elements of $\mathcal{A}$, requires fixing a basis for representation. It is more convenient to choose a basis in which all $H_i$s are diagonal:
\begin{equation}
H_i|\lambda_1,\cdots,\lambda_l\rangle=\lambda_i|\lambda_1,\cdots\lambda_l\rangle.
\end{equation}
Vector $|\lambda\rangle$ with elements $\lambda_i$ is called weight vector. Using commutation relation in \eqref{HE}, it is easy to see that vectors 
$E_{\pm\alpha}|\lambda_1,\cdots,\lambda_l\rangle$ are eigenvectors of $H_i$ with weight $\lambda_i\pm\alpha_i$. Hence $E_{\pm\alpha}$ are raising and lowering operators. The highest weight denoted by $|\Lambda\rangle$ is unique and for any positive root $\alpha$, we have $E_{\alpha}|\Lambda\rangle=0$. This highest weight, determines the dimension of irreducible representation of the algebra $\mathcal{A}$. It is helpful to remind that $su(2)$ algebra, is rank one where $J_z$ performs as $H_1$ and ladder operators are $J_{\pm}$. We also know that the highest weight of representation $|m=j\rangle$ corresponds to $2j+1$ dimensional irreducible representation for $su(2)$.\\

Going back to non unitary dynamics of quantum systems, if it is a Markovian process described by Lindblad operators  $\{E_{\alpha}\}$ which are ladder operators of a semi-simple algebra $\mathcal{A}$, that is
\begin{equation}
\mathcal{L}(\rho)=\sum_{\alpha}2E_{\alpha}\rho E_{\alpha}^{\dag}-\{\ E_{\alpha}^{\dag}E_{\alpha},\rho\},
\end{equation}
then the Casimir operator of the algebra $C$ is constant of motion because by definition it commutes with all elements of algebra (in $su(2)$ the Casimir operator is $J^2$). If the initial state of the system is given by common eigenvector of Casimir operator $C$ and all basis of Cartan subalgebra $|c,\lambda_1,\cdots,\lambda_l\rangle$, that is 
\begin{eqnarray}
&&C|c,\lambda_1,\cdots,\lambda_l\rangle=c|c,\lambda_1,\cdots,\lambda_l\rangle,\cr\cr
&&H_i|c,\lambda_1,\cdots,\lambda_l\rangle= \lambda_i |c,\lambda_1,\cdots,\lambda_l\rangle,
\end{eqnarray}
during the evolution system remains in a subspace where Casimir operator has value $c$. Actually, the Lindblad operators just change the weights and do not lead the dynamics out of a subspace specified by the value of Casimir operator. Dimension of this subspace is determined by the highest weight corresponding to the initial state, which is given by $(E_{\alpha})^q |c,\lambda_1,\cdots,\lambda_l\rangle$, where $q$ is the largest positive integer that $(E_{\alpha})^q |c,\lambda_1,\cdots,\lambda_l\rangle\neq 0$ and $\alpha$ is a positive root. \\
\section{Explicit form of eigenvectors of matrix $\mathcal{M}$} \label{AppB}
Here we represent the explicit form of eigenvectors of $(3N-1)\times(3N-1)$ matrix $\mathcal{M}$, which is defined in equation \eqref{incoherent-M-def}. We write the matrix form of $\mathcal{M}$ in the same basis as the vector \eqref{incoherent-coef-vec}, that is the basis $\{ |j\rangle, j=1..3N-1 \}$, in which the elements of $\mathcal{M}$ can be written as follows:\\
For $1\leq j \leq N$:
\begin{eqnarray*}
\mathcal{M}_{ij}&=&tr(\Lambda_i \mathcal{L}[\Lambda_j]) \hspace{2.2cm} 1\leq i \leq N \cr\cr
\mathcal{M}_{ij}&=&tr(\Omega_{i-N-1} \mathcal{L}[\Lambda_j]) \hspace{0.6cm} N+1 \leq i \leq 2N \cr\cr
\mathcal{M}_{ij}&=&tr(\chi_{i-2N} \mathcal{L}[\Lambda_j]) \hspace{0.7 cm} 2N+1 \leq i \leq 3N-1.\cr\cr
\end{eqnarray*}
For $N+1\leq j\leq 2N$:
\begin{eqnarray*}
\mathcal{M}_{ij}&=&tr(\Lambda_i \mathcal{L}[\Omega_{N-j-1}]) \hspace{2.2cm} 1\leq i \leq N \cr\cr
\mathcal{M}_{ij}&=&tr(\Omega_{i-N-1} \mathcal{L}[\Omega_{N-j-1}]) \hspace{0.6cm} N+1 \leq i \leq 2N \cr\cr
\mathcal{M}_{ij}&=&tr(\chi_{i-2N} \mathcal{L}[\Omega_{N-j-1}]) \hspace{0.7 cm} 2N+1 \leq i \leq 3N-1.\cr\cr
\end{eqnarray*}
And for $2N+1\leq j \leq 3N-1$:
\begin{eqnarray*}
\mathcal{M}_{ij}&=&tr(\Lambda_i \mathcal{L}[\chi_{j-2N}]) \hspace{2.2cm} 1\leq i \leq N \cr\cr
\mathcal{M}_{ij}&=&tr(\Omega_{i-N-1} \mathcal{L}[\chi_{j-2N}]) \hspace{0.6cm} N+1 \leq i \leq 2N \cr\cr
\mathcal{M}_{ij}&=&tr(\chi_{i-2N} \mathcal{L}[\chi_{j-2N}]) \hspace{0.7 cm} 2N+1 \leq i \leq 3N-1.\cr\cr
\end{eqnarray*}
Using this notation the explicit form of eigenvectors with zero eigenvalue is as follows:
\begin{equation}
|r_1^{(0)}\rangle = \frac{1-\nu}{(1+\nu)(1-\nu^N)}  \sum_{j=1}^{N} \left(\nu^j |j \rangle + \nu^{j-1} |N+j \rangle \right) ,
\end{equation}
with the corresponding left eigenvector:
\begin{equation}
\begin{aligned}
\langle l_1^{(0)} |=\frac{(1+\nu)(1-\nu^N)}{N (1 -\nu^{N+1})} &\big[ \sum_{j=1}^{N}( j \langle j | + (N-j+1) \langle N+j |)
\\
&+ 2  \sum_{j=1}^{N-1}  \sqrt{j(N-j)}  \langle 2N+j | \big],
\end{aligned}
\end{equation}
and 
\begin{equation}
\begin{aligned}
|r_2^{(0)}\rangle = \frac{1-\nu}{N (\nu -\nu^N)} &\big[ \sum_{j=1}^{N} (N-j) \nu^j |j \rangle + (j-1) \nu^{j-1} |N+j \rangle
\\
& - \sum_{j=1}^{N-1}  \sqrt{j(N-j)} \nu^j |2N+j \rangle  \big] \cr
\end{aligned}
\end{equation}
with the left eigenvector 
\begin{eqnarray}
\langle l_2^{(0)} |&=&\frac{(\nu-\nu^N)}{N (1 -\nu^{N+1})} ( -N  \langle N+1|+\sum_{j=1}^{2N-1}(\alpha|N-j|-N)\langle j|\cr\cr
&-&2 \alpha \sum_{j=1}^{N-1}  \sqrt{j(N-j)} \langle 2N+j |  ).
\end{eqnarray}
Here
$\alpha=\frac{(1-\nu^N)(1+\nu)}{\nu-\nu^N}$. Normalization factors are chosen such that density operator corresponding to each right eigenvalue has trance one and also $\langle l_i ^{(0)} | r_j^{(0)}\rangle =\delta_{ij}$.\\
\section{Explicit form of eigenvectors of matrix $\mathcal{M'}$} \label{AppC}
To find the eigenvectors corresponding to zero eigenvalues of matrix $\mathcal{M}'$ (its elements are given in equation \eqref{Mprime-Elements}) again we use the notation $|j\rangle$ to represent $(N+1)$ basis vectors of equation \eqref{coherent-coef-vec}. $\mathcal{M'}$ has only one zero eigenvalue and  we find its corresponding right and left eigenvectors to be:
\begin{equation}
\begin{aligned}
|{r'}^{(0)}\rangle &= \frac{1-\nu}{1-\nu^{N+1}} \sum_{j=0}^{N} \nu^j |j\rangle
\\
\langle {l'}^{(0)}| &= \sum_{n=0}^{N} \langle j| 
\end{aligned}
\end{equation}


{}
\end{document}